\documentclass[smallextended]{svjour3}
\usepackage{graphicx, amsmath, amssymb}
\usepackage[caption=false]{subfig}

\newcommand{\ket}[1]{{\left| #1 \right\rangle}}
\newcommand{\ketbra}[2]{{\left| #1 \middle\rangle \middle \langle #2 \right|}}
\newcommand{\fref}[1]{Fig.~\ref{#1}}

\journalname{Quantum Inf Process}

\begin{document}

\title{Search on Vertex-Transitive Graphs by Lackadaisical Quantum Walk}

\author{Mason L.~Rhodes \and Thomas G.~Wong}

\authorrunning{M.~L.~Rhodes \and T.~G.~Wong}

\institute{Department of Physics, Creighton University, 2500 California Plaza, Omaha, Nebraska 68178, USA \\
	\email{masonrhodes@creighton.edu} \\
	\email{thomaswong@creighton.edu}
}

\date{Received: date / Accepted: date}

\maketitle

\begin{abstract}
	The lackadaisical quantum walk is a discrete-time, coined quantum walk on a graph with a weighted self-loop at each vertex. It uses a generalized Grover coin and the flip-flop shift, which makes it equivalent to Szegedy's quantum Markov chain. It has been shown that a lackadaisical quantum walk can improve spatial search on the complete graph, discrete torus, cycle, and regular complete bipartite graph. In this paper, we observe that these are all vertex-transitive graphs, and when there is a unique marked vertex, the optimal weight of the self-loop equals the degree of the loopless graph divided by the total number of vertices. We propose that this holds for all vertex-transitive graphs with a unique marked vertex. We present a number of numerical simulations supporting this hypothesis, including search on periodic cubic lattices of arbitrary dimension, strongly regular graphs, Johnson graphs, and the hypercube.
\end{abstract}


\section{Introduction}

The coined quantum walk is a quantum analogue of the discrete-time random walk. It was first discovered by Meyer \cite{Meyer1996a} in the context of quantum cellular automata, and he proved \cite{Meyer1996b} that an internal degree of freedom was necessary to make the evolution nontrivial. He also proved that the coined quantum walk is a discretization of the Dirac equation of relativistic quantum mechanics, with the internal degree of freedom playing the role of spin. Later, Aharonov \textit{et al.}~\cite{Aharonov2001} and Ambainis \textit{et al.}~\cite{Ambainis2001} reframed this in the context of quantum walks, with the internal degree of freedom playing the role of a coin that specifies the direction of a walker. Since then, coined quantum walks have led to the discovery of several quantum algorithms, including algorithms for searching \cite{SKW2003}, solving element distinctness \cite{Ambainis2004}, finding triangles in graphs or networks \cite{MSS2005}, and evaluating boolean formulas \cite{FGG2008}.

The walk is encoded on a graph of $N$ vertices, where the vertices specify the possible positions of the walker, and the edges specify the directions along which the walker can face. If the graph is regular with degree $d$, the Hilbert space of the quantum walk is $\mathbb{C}^N \otimes \mathbb{C}^d$, which is the tensor product of the vertex space and the coin space. One step of a coined quantum walk consists of applying a quantum coin flip $(I_N \otimes C)$ followed by a shift $S$ to adjacent vertices, so $U_\text{walk} = S(I_N \otimes C)$.

To search for a marked vertex using a quantum walk, the walker $\ket{\psi(t)}$ begins in a uniform superposition over the vertices and directions:
\begin{equation}
	\label{eq:psi0}
	\ket{\psi(0)} = \ket{s_v} \otimes \ket{s_c},
\end{equation}
where
\[ \ket{s_v} = \frac{1}{\sqrt{N}} \sum_{i=1}^N \ket{i}, \quad \ket{s_c} = \frac{1}{\sqrt{d}} \sum_{i=1}^d \ket{i}. \]
This expresses our initial lack of information as to which vertex is marked, so we guess them all equally. It can also be prepared without knowledge of the marked vertex.

From this initial state, we alternate between querying an oracle and taking a step of the quantum walk. We use the phase-flip oracle $Q$ that negates the amplitude at the marked vertex \cite{AKR2005}. In Dirac notation, if the marked vertex is $\ket{w}$, then $Q = (I_N - 2 \ketbra{w}{w})$ is applied to the vertex space, and the identity $I_d$ is applied to the coin space. For the coin flip, we use the Grover diffusion coin \cite{SKW2003}, which inverts the coin amplitudes at a vertex about their average at that vertex. That is, we apply the identity $I_N$ to the vertex space and
\begin{equation}
	\label{eq:C}
	C = 2 \ketbra{s_c}{s_c} - I_d
\end{equation}
to the coin space. Finally, we use the flip-flop shift \cite{AKR2005}, which causes the particle to hop and turn around. So if $\ket{uv}$ denotes a particle at vertex $u$ pointing to vertex $v$, then $S\ket{uv} = \ket{vu}$. Altogether, the search is performed by repeatedly applying
\begin{equation}
	\label{eq:U}
	U = S(I_N \otimes C)(Q \otimes I_d).
\end{equation}
This search operator is applied some number of steps, i.e., the runtime of the algorithm, so that amplitude accumulates at the marked vertex, and then the position of the walker is measured. With some probability, the walker will be at the marked vertex. Otherwise, the algorithm may have to be repeated.

This search algorithm has been applied to a large variety of graphs, such as the hypercube \cite{SKW2003}, complete graph \cite{AKR2005}, cubic lattices of arbitrary dimension \cite{AKR2005}, complete bipartite graphs \cite{Reitzner2009,Wong31}, strongly regular graphs \cite{Xue2017}, and tetrahedral graphs \cite{Xue2019}.

\begin{figure*}
\begin{center}
        \subfloat[] {
		\includegraphics[height=1.4in]{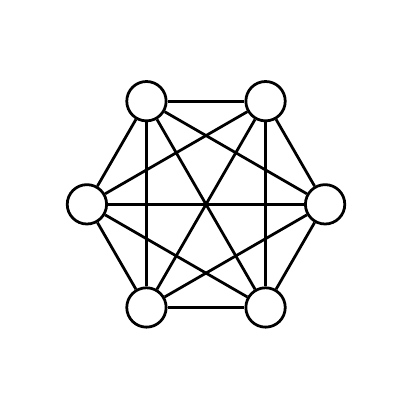}
                \label{fig:complete}
        } \quad
        \subfloat[] {
		\includegraphics[height=1.4in]{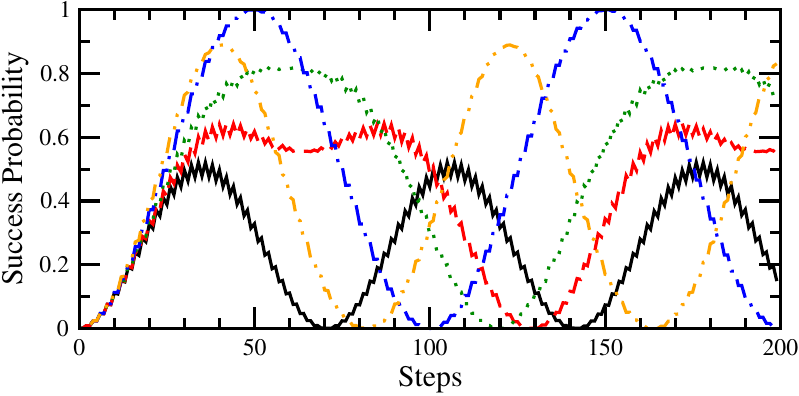}
                \label{fig:complete_1024}
        }

        \subfloat[] {
		\includegraphics[height=1.4in]{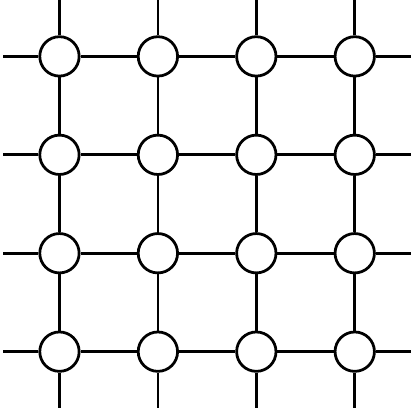}
		\label{fig:torus}
        } \quad
        \subfloat[] {
		\includegraphics[height=1.4in]{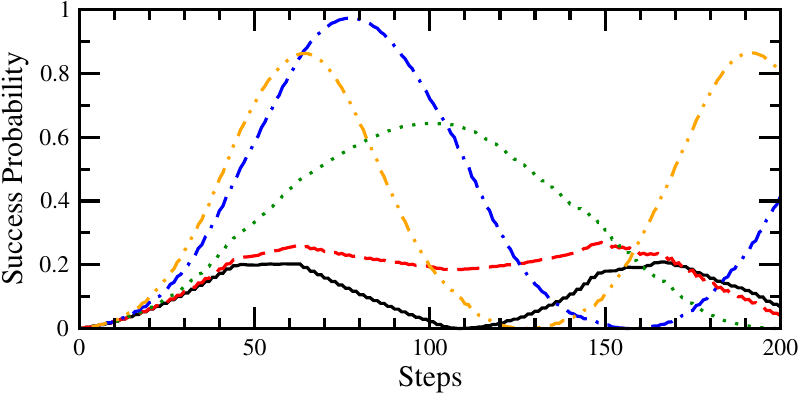}
		\label{fig:torus_1024}
        }

        \subfloat[] {
                \includegraphics[height=1.4in]{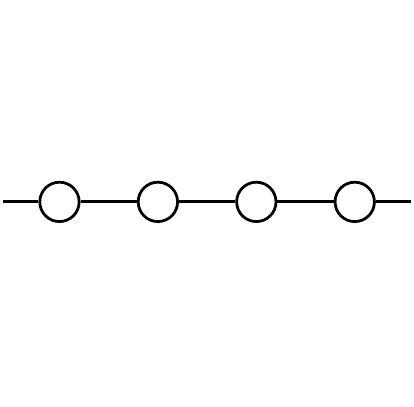}
		\label{fig:cycle}
        } \quad
        \subfloat[] {
		\includegraphics[height=1.4in]{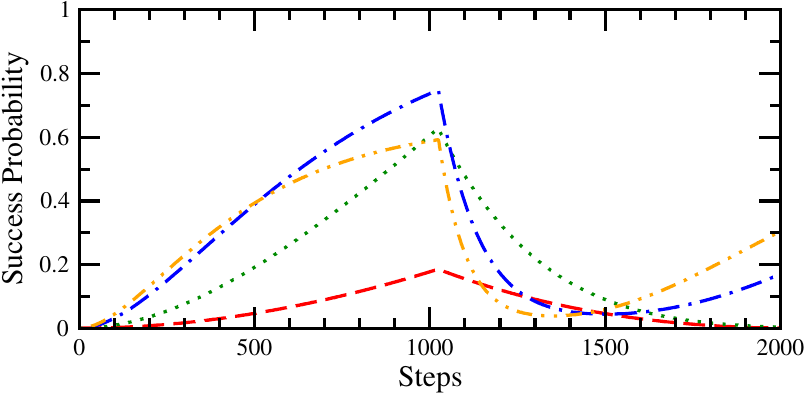}
		\label{fig:cycle_1024}
        }

        \subfloat[] {
                \includegraphics[height=1.4in]{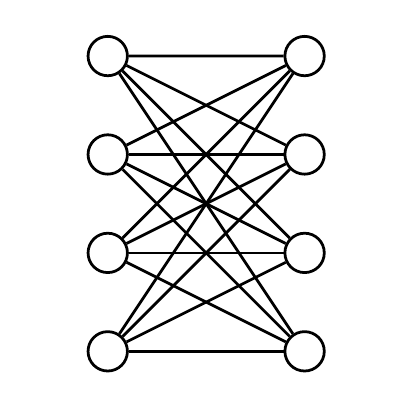}
                \label{fig:bipartite}
        } \quad
        \subfloat[] {
                \includegraphics[height=1.4in]{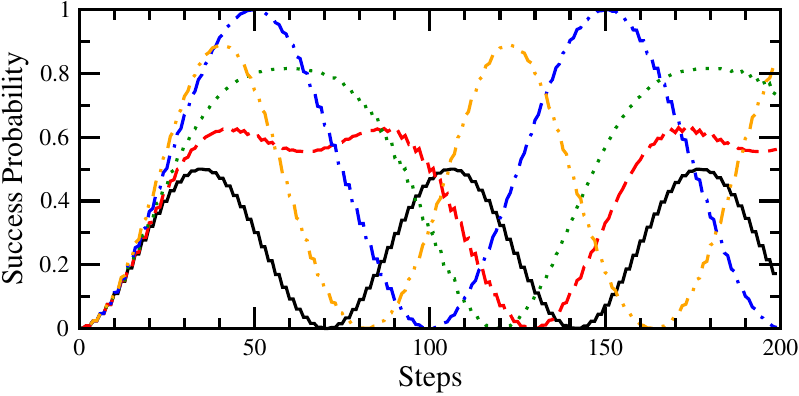}
                \label{fig:bipartite_1024}
        }

	\caption{Previous results for search by lackadaisical quantum walk on vertex-transitive graphs with a unique marked vertex. (a) A complete graph of $N = 6$ vertices. (b) Lackadaisical quantum search on the complete graph of $1024$ vertices. The solid black curve is $\ell = 0$, dashed red is $0.2$, dotted green is $0.4$, the dot-dashed blue is $1$, and the dot-dot-dashed orange is $2$. (c) A torus, or periodic square lattice, with $16$ vertices. (d) Lackadaisical quantum search on the torus with $1024$ vertices. The solid black curve is $\ell = 0$, dashed red is $0.0002$, dotted green is $0.001$, the dot-dashed blue is $0.0039$, and the dot-dot-dashed orange is $0.008$. (e) A cycle, or one-dimensional lattice with periodic boundary conditions, with $4$ vertices. (f) Lackadaisical quantum search on the cycle with $1024$ vertices. The solid black curve is $\ell = 0$ and is a horizontal line at $1/1024$ that is covered by the $x$-axis, dashed red is $0.0001$, dotted green is $0.0005$, the dot-dashed blue is $0.0019531$, and the dot-dot-dashed orange is $0.003$. (g) A regular complete bipartite graph of $N = 8$ vertices. (h) Lackadaisical quantum search on the regular complete bipartite graph of $1024$ vertices. The solid black curve is $\ell = 0$, dashed red is $0.1$, dotted green is $0.2$, the dot-dashed blue is $0.5$, and the dot-dot-dashed orange is $1$.} 
\end{center}
\end{figure*}

Typical quantum walks occur on simple graphs, where there are no self-loops. Ambainis, Kempe, and Rivosh \cite{AKR2005}, however, explored searching the complete graph with one self-loop added to each vertex. The complete graph with $N = 6$ vertices is depicted in \fref{fig:complete}, and since all vertices are adjacent to each other, it corresponds to the unstructured database of Grover's algorithm \cite{Grover1996}. With a self-loop at each vertex, the position and coin spaces both have dimension $N$, and two steps of $U$ \eqref{eq:U} were proved to be equivalent to one step of Grover's algorithm. As a result of including a self-loop at each vertex, the maximum success probability of the algorithm increased from its loopless value of $1/2$ to $1$, for large $N$.

Wong \cite{Wong10} generalized this idea by putting $\ell$ integer self-loops on each vertex. Then the number of self-loops is a parameter that adjusts the likelihood of a walker staying put. This forms a type of lazy quantum walk, called a lackadaisical quantum walk. Later, he updated the definition by using a single self-loop of weight $\ell$ at each vertex. Then if $\ell$ is in integer, it is equivalent to having $\ell$ unweighted self-loops in his original formulation, but now $\ell$ can take non-integer values as well \cite{Wong27}. On the line, this is similar to the three-state quantum walk \cite{Inui2005}.

Mathematically, the weighted self-loop is accounted for by modifying $\ket{s_c}$, which in turn changes the initial state \eqref{eq:psi0} and Grover coin \eqref{eq:C}. Keeping $d$ as the degree of the loopless graph, the coin space is now $(d+1)$-dimensional, and then
\begin{equation}
	\label{eq:sc}
	\ket{s_c}  = \frac{1}{\sqrt{d + \ell}} \left( \sqrt{\ell} \ket{\circlearrowleft} + \sum_{i=1}^d \ket{i} \right).
\end{equation}
That is, the amplitude along the self-loop is weighted by $\sqrt{\ell}$, while the amplitudes pointing to the other vertices remain unweighted. We also change $I_d$ to $I_{d+1}$ in \eqref{eq:C} and \eqref{eq:U}, and the lackadaisical quantum walk is simply the coined quantum walk with these changes. Note \eqref{eq:sc} is defined this way so that the coined quantum walk is equivalent to Szegedy's quantum Markov chain \cite{Wong27,Szegedy2004}.

Search using the lackadaisical quantum walk has been explored on the complete graph \cite{Wong27}, discrete torus with one marked vertex \cite{Wong28} and multiple marked vertices \cite{Saha2018,Nahimovs2019,Giri2019}, cycle \cite{Giri2019}, and complete bipartite graph \cite{Wong32}. Depending on the weight of the self-loop $\ell$, the success probability can be improved. In the next section, we review these results when the graphs are vertex-transitive and have a unique marked vertex. From this, we observe a trend that $\ell = d/N$ is optimal in all these cases. We hypothesize that this is true in general for vertex-transitive graphs with unique marked vertices, and in the following sections, we give additional numerical evidence for this trend. We simulate search by lackadaisical quantum walk on periodic cubic lattices of arbitrary dimension, strongly regular graphs, Johnson graphs, and the hypercube. We end with remarks on proofs of these observations.


\section{Observation from Previous Results}

As described in the introduction, the lackadaisical quantum walk was introduced in the context of searching the complete graph, which is the quantum-walk formulation of Grover's algorithm. Wong analytically proved that $\ell = 1$ maximizes the success probability when searching the complete graph \cite{Wong10,Wong27}. This is demonstrated numerically in \fref{fig:complete_1024} for the complete graph of $N = 1024$ vertices. The success probability starts at $1/1024 = 0.00097656$. As $U$ \eqref{eq:U} is applied, the success probability increases. When $\ell = 0$, the solid black curve reaches a peak success probability of $1/2$ at $\pi \sqrt{N}/2\sqrt{2} = 36$ steps. At this time, the position of the particle is measured, and the marked vertex is found with probability $1/2$, so on average, the algorithm must be repeated twice before a marked vertex is found. When $\ell = 0.2$, the dashed red curve indicates that the success probability reaches a higher value. Further increasing the weight to $\ell = 0.4$, the green dotted curve reaches a greater success probability still. When $\ell = 1$, the dot-dashed blue curve reaches a success probability of 1. Further increasing $\ell$, such as $\ell = 2$ in the dot-dot-dashed orange curve, results in a smaller peak success probability. So there is some optimal amount of laziness that boosts the success probability the most. We note that the optimal weight for this, $\ell = 1$, is asymptotically the degree of the loopless graph divided by the number of vertices, i.e., $d/N = (N-1)/N \approx 1$. This optimality of $\ell = 1$ holds with multiple marked vertices as well \cite{Wong10}.

Historically, the next graph considered for search by lackadaisical quantum walk was the torus, or 2D square lattice with periodic boundary conditions \cite{Wong28}. An example with $N = 16$ vertices is depicted in \fref{fig:torus}. Wong numerically determined that the optimal weight self-loop was $\ell = 4/N$. For example, this can be seen in \fref{fig:torus_1024}, which shows search on the torus with $N = 1024$ vertices. The solid black curve corresponds to $\ell = 0$, and the peak success probability reaches around 0.2, which scales with $N$ as $O(1/\log N)$ \cite{AKR2005}. As $\ell$ increases, however, the peak success probability increases, eventually reaching roughly 0.97 when $\ell = 4/N = 0.0039$, as shown by the dot-dashed blue curve. With $\ell = 4/N$, the success probability is constant in $N$, i.e., $O(1)$. Thus, the lackadaisical quantum walk improved the success probability from $O(1/\log N)$ to $O(1)$. As before, increasing $\ell$ beyond its optimal value decreases the peak success probability. We note that $\ell = 4/N$ is the degree of the loopless graph divided by the total number of vertices, i.e., $d/N$, since $d = 4$. With multiple marked vertices \cite{Saha2018,Nahimovs2019,Giri2019}, it is no longer true that $d/N$ is the best value of $\ell$.

Next, the cycle was explored by Giri and Korepin \cite{Giri2019}. The cycle is a 1D periodic lattice; an example with $N = 4$ vertices is shown in \fref{fig:cycle}. They numerically showed that a significant improvement in the success probability occurs when $\ell = 2/N$. When $\ell = 0$, the success probability stays at its initial value of $1/N$, no matter how many steps are taken \cite{Wong25}. When $\ell = 2/N$, it reaches a constant value $O(1)$. This is demonstrated in \fref{fig:cycle_1024} for search on the cycle of $N = 1024$ vertices. When $\ell = 0$, the success probability stays at its initial value of $1/1024$. The solid black curve corresponding to this, however, is hidden by the $x$-axis. As $\ell$ increases, the success probability increases. When $\ell = 2/N = 0.0019531$, as shown by the dot-dashed blue curve, the success probability reaches 0.747. As $\ell$ is further increased, the success probability drops. We note that $\ell = 2/N$ is the degree of the loopless graph divided by the total number of vertices, i.e., $d/N$, since $d = 2$. We also note that $\ell = 1/N$ results in a slightly higher peak success probability, but it is still constant, so there is no additional improvement to the runtime scaling. That is, $\ell = 2/N$ is sufficient to improve the success probability from $1/N$ to $O(1)$.

Most recently, search on the complete bipartite graph was explored \cite{Wong32}. An example with $N = 8$ vertices is shown in \fref{fig:bipartite}. When the graph is regular and there is a unique marked vertex, Rhodes and Wong \cite{Wong32} analytically proved that $\ell = 1/2$ boosts the success probability to 1. This is demonstrated in \fref{fig:bipartite_1024} with $N = 1024$ vertices. When $\ell = 0$, the solid black curve shows that the success probability reaches a maximum value of $1/2$. As $\ell$ increases, the success probability increases, reaching a value of $1$ when $\ell = 1/2$, as depicted by the dot-dashed blue curve. As expected, increasing $\ell$ beyond this lowers the success probability. We note that $\ell = 1/2$ is the degree of the loopless graph divided by the total number of vertices, i.e, $d/N$, since $d = N/2$. When there are multiple marked vertices or the graph is irregular, the optimal value of $\ell$ is no longer $1/2$.

We observe that in each of these previous works, when the graphs were vertex transitive and contained a unique marked vertex, the optimal value of $\ell$ was $d/N$. In the next several sections, we provide several examples of other vertex-transitive graphs where search for a unique marked vertex supports this observation. This suggests that $\ell = d/N$ is optimal in general for vertex-transitive graphs with unique marked vertices.

A similar observation regarding the optimal value of $\ell$ was given by Wang \textit{et al.} \cite{Wang2017_2D} in 2017. At the time, only the complete graph and torus had been considered for search by lackadaisical quantum walk. They speculated that the ``proper weight of each self-loop may be equal to the degree centrality of its corresponding vertex,'' i.e., the degree of the vertex divided by $(N-1)$. More recent results have disproved this speculation, however. Search on the torus with multiple marked vertices \cite{Saha2018,Nahimovs2019,Giri2019}, and search on the complete bipartite graph with one or more marked vertices \cite{Wong32}, indicate that $\ell$ should not be the degree centrality at each vertex. In this paper, our observation is specifically restricted to vertex-transitive graphs with a single marked vertex, where the observation does seem to hold.


\section{Arbitrary-Dimensional Cubic Lattices}

\begin{figure}
\begin{center}
        \subfloat[] {
		\includegraphics[width=1.4in]{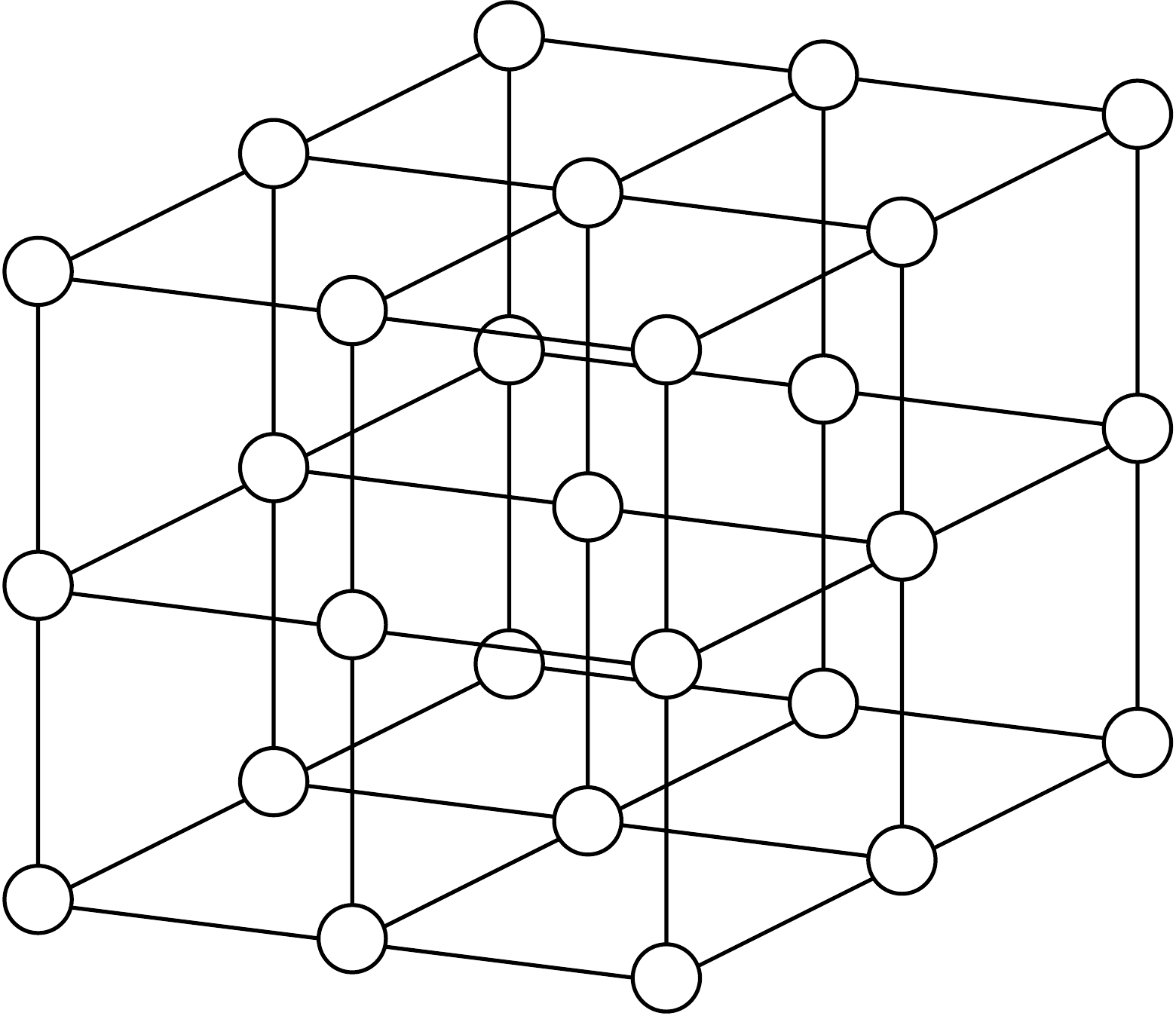}
                \label{fig:cubic}
        } \quad
        \subfloat[] {
                \includegraphics[height=1.4in]{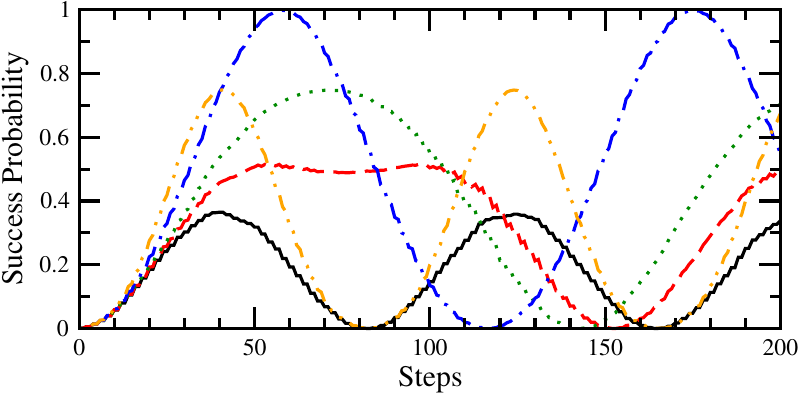}
                \label{fig:cubic_3_10}
        }

	\subfloat[] {
                \includegraphics[height=1.4in]{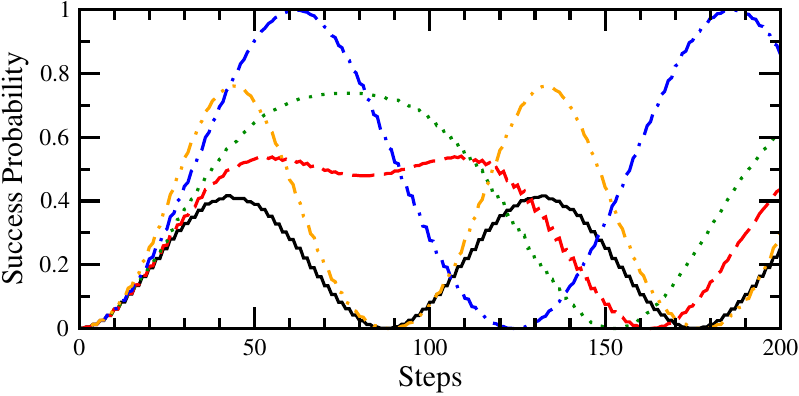}
                \label{fig:cubic_4_6}
        }

	\subfloat[] {
                \includegraphics[height=1.4in]{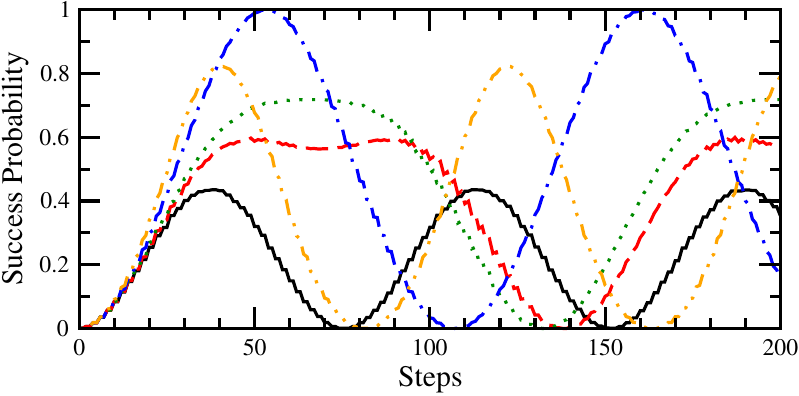}
                \label{fig:cubic_5_4}
        }

	\caption{(a) A cubic lattice of $27$ vertices. The boundaries are periodic, so they wrap around in each dimension. (b) Lackadaisical quantum search on the 3D lattice of 1000 vertices. The solid black curve is $\ell = 0$, dashed red is $0.001$, dotted green is $0.002$, the dot-dashed blue is $0.006$, and the dot-dot-dashed orange is $0.018$. (c) Lackadaisical quantum search on the 4D lattice of 1296 vertices. The solid black curve is $\ell = 0$, dashed red is $0.001$, dotted green is $0.002$, the dot-dashed blue is $0.006173$, and the dot-dot-dashed orange is $0.018$. (d) Lackadaisical quantum search on the 5D lattice of 1024 vertices. The solid black curve is $\ell = 0$, dashed red is $0.002$, dotted green is $0.003$, the dot-dashed blue is $0.009766$, and the dot-dot-dashed orange is $0.024$.} 
\end{center}
\end{figure}

Since $\ell = d/N$ is optimal for both the cycle and torus, which are 1D and 2D periodic lattices, a reasonable extension is to explore search on arbitrary-dimensional periodic cubic lattices. A 3D example with $N = 27$ vertices is shown in \fref{fig:cubic}. For a $D$-dimensional lattice, the degree is $d = 2D$, and our numerical simulations support that $\ell = d/N = 2D/N$ is optimal. For example, search on the 3D periodic cubic lattices with $N = 1000$ vertices is shown in \fref{fig:cubic_3_10}, the dot-dashed blue curve shows that the success probability reaches $1$ when $\ell = 6/1000 = 0.006$. Similarly, search on the 4D periodic cubic lattice with $N = 1296$ vertices is shown in \fref{fig:cubic_4_6}, the dot-dashed blue curve shows that the success probability reaches $1$ when $\ell = 8/1296 = 0.006173$. Finally, search on the 5D periodic lattice with $N = 1024$ is shown in \fref{fig:cubic_5_4}, the dot-dashed blue curve shows that the success probability reaches $1$ when $\ell = 10/1024 = 0.009766$. We simulated up to 10D lattices, and the hypothesis of the optimality of $\ell = d/N$ held in all cases.


\section{Strongly Regular Graphs}

\begin{figure*}
\begin{center}
        \subfloat[] {
                \includegraphics[height=1.4in]{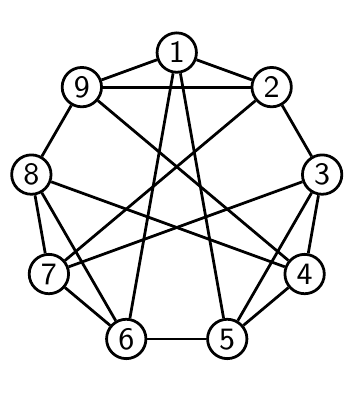}
		\label{fig:srg_paley}
        } \quad
        \subfloat[] {
		\includegraphics[height=1.4in]{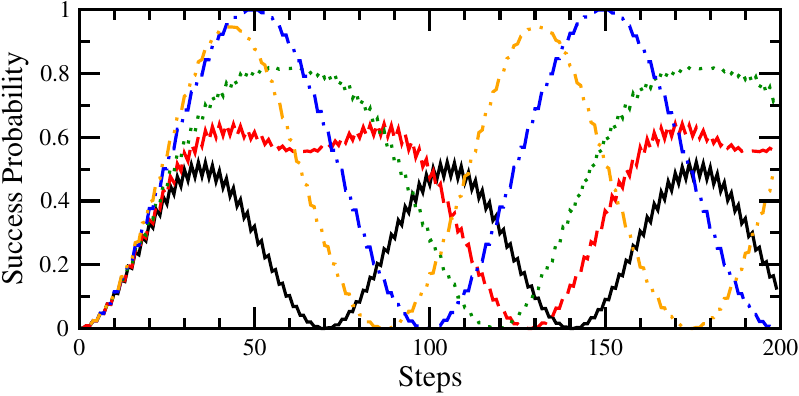}
		\label{fig:srg_1009_504_251_252}
        }

        \subfloat[] {
                \includegraphics[height=1.4in]{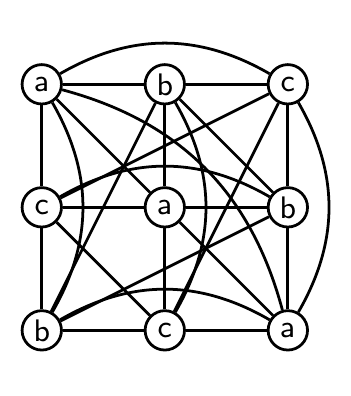}
		\label{fig:srg_latin}
        } \quad
        \subfloat[] {
		\includegraphics[height=1.4in]{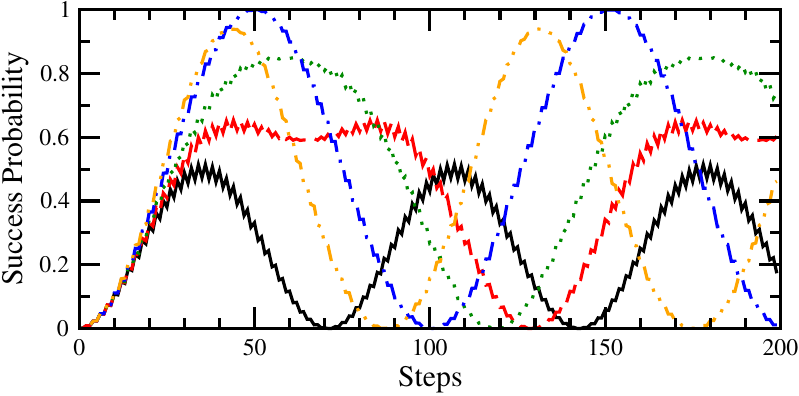}
		\label{fig:srg_1024_93_32_6}
        }

        \subfloat[] {
                \includegraphics[height=1.4in]{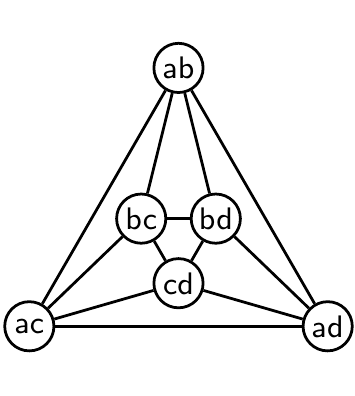}
		\label{fig:srg_triangular}
        } \quad
        \subfloat[] {
		\includegraphics[height=1.4in]{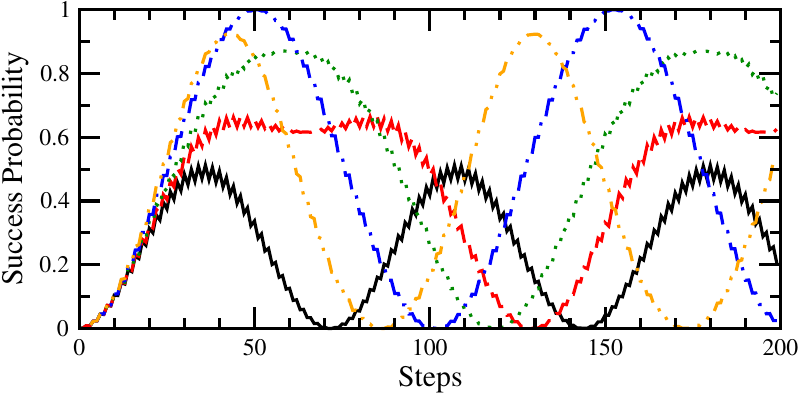}
		\label{fig:srg_1035_88_44_4}
        }

	\caption{(a) Paley graph with strongly regular parameters $(9,4,1,2)$. (b) Lackadaisical quantum search on the Paley graph with strongly regular parameters $(1009,504,251,252)$. The solid black curve is $\ell = 0$, dashed red is $0.1$, dotted green is $0.2$, the dot-dashed blue is $0.499504$, and the dot-dot-dashed orange is $0.8$. (c) Latin square graph with strongly regular parameters $(9,6,3,6)$. (d) Lackadaisical quantum search on the Latin square graph with strongly regular parameters $(1024,93,32,6)$. The solid black curve is $\ell = 0$, dashed red is $0.02$, dotted green is $0.04$, the dot-dashed blue is $0.090820$, and the dot-dot-dashed orange is $0.15$. (e) Triangular graph $T_{4}$ with strongly regular parameters $(6,4,2,4)$. (f) Lackadaisical quantum search on the Triangular graph $T_{46}$ with strongly regular parameters $(1035,88,44,4)$. The solid black curve is $\ell = 0$, dashed red is $0.02$, dotted green is $0.04$, the dot-dashed blue is $0.0850242$, and the dot-dot-dashed orange is $0.15$. } 
\end{center}
\end{figure*}

The regular complete bipartite graph in \fref{fig:bipartite} is an example of a strongly regular graph. A strongly regular graph has parameters $(N,d,\lambda,\mu)$. First, it is regular with $N$ vertices and degree $d$. Furthermore, to be strongly regular, all pairs of adjacent vertices have $\lambda$ mutual neighbors, and all pairs of non-adjacent vertices have $\mu$ mutual neighbors. The regular complete bipartite graph has parameters $(N, N/2, 0, N/2)$. To elaborate, it has $N$ vertices, and each is adjacent to half the vertices, so $d = N/2$. Adjacent vertices must be in opposite partite sets, so they have zero mutual neighbors, so $\lambda = 0$. Nonadjacent vertices are in the same partite set, so they are mutually adjacent to all $N/2$ vertices in the other partite set, so $\mu = N/2$. In this section, we explore whether other strongly regular graphs follow our observation that the optimal weight is $\ell = d/N$.
 
Although not all strongly regular graphs are known, several families exist. Paley graphs are one such family. Their vertices are labeled $0, 1, \dots, N-1$, where $N$ is a prime power. Vertices are adjacent if and only if their difference is a quadratic residue. From Chapter 10 of \cite{Godsil2001}, they are strongly regular with parameters
\[ N, \quad k = \frac{N-1}{2}, \quad \lambda = \frac{N-5}{4}, \quad \mu = \frac{N-1}{4}. \]
For example, Figure~\ref{fig:srg_paley} depicts the Paley graph $(9,4,1,2)$. It has 9 vertices, each with 4 neighbors. Any pair of adjacent vertices has 1 mutual neighbor, and any pair of nonadjacent vertices has 2 mutual neighbors. In Figure~\ref{fig:srg_1009_504_251_252}, we show search on the Paley graph $(1009,504,251,252)$ by lackadaisical quantum walk. From the dot-dashed blue curve, when $\ell = d/N = 504/1009 = 0.499504$, the success probability reaches 1, supporting the proposal that $\ell = d/N$ is best.

Another family of strongly regular graphs are Latin square graphs. An example with $9$ vertices is shown in \fref{fig:srg_latin}, and it has parameters $(9,6,3,6)$. Latin square graphs can be drawn as $\sqrt{N} \times \sqrt{N}$ square lattices. First, vertices are adjacent if they are in the same row or same column. Furthermore, $\sqrt{N}$ symbols appear exactly once in every row and column, and vertices are also adjacent if they share the same symbol. In \fref{fig:srg_latin}, the symbols are $a$, $b$, $c$. Latin square graphs are strongly regular with parameters \cite{Cameron1991}
\[ N, \quad k = 3(\sqrt{N} - 1), \quad \lambda = \sqrt{N}, \quad \mu = 6. \]
In \fref{fig:srg_latin}, search by lackadaisical quantum walk on the Latin square graph $(1024,93,32,6)$ is shown. The dot-dashed blue curve indicates that the success probability is boosted to $1$ when $\ell = d/N = 93/1024 = 0.090820$, consistent with the observation.

As a final family of strongly regular graphs to investigate, the triangular graph $T_m$ is the line graph of the complete graph of $m$ vertices, where $m \ge 4$. For example, $T_4$ is shown in \fref{fig:srg_triangular}. Another way to construct it is by labeling the vertices with the two-element subsets of four symbols, and vertices are adjacent if they differ in exactly one symbol. In \fref{fig:srg_triangular}, the symbols are $a,b,c,d$, so the two-element subsets are $ab$, $ac$, $ad$, $bc$, $bd$, and $cd$. Triangular graphs are strongly regular with parameters \cite{Cameron1991}
\[ N = \frac{m(m-1)}{2}, \enspace k = 2(m-2), \enspace \lambda = m-2, \enspace \mu = 4. \]
So $T_4$ has parameters (6,4,2,4). In \fref{fig:srg_1035_88_44_4}, we plot the success probability for search on $T_{46}$, which has strongly regular parameters $(1035,88,44,4)$. As expected, the dot-dashed blue curve shows that the success probability reaches $1$ when $\ell = d/N = 88/1035 = 0.0850242$.


\section{Johnson Graphs}

\begin{figure*}
\begin{center}
        \subfloat[] {
		\includegraphics[height=1.4in]{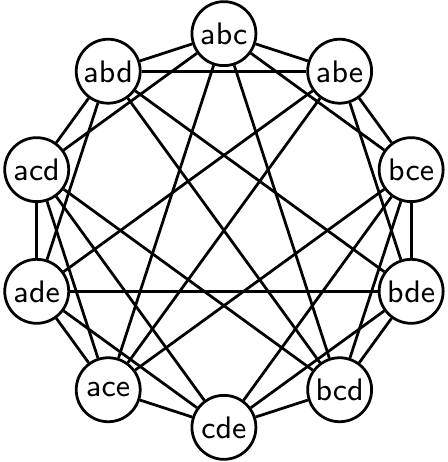}
                \label{fig:johnson_graph}
        } \quad \quad
	\subfloat[] {
                \includegraphics[height=1.4in]{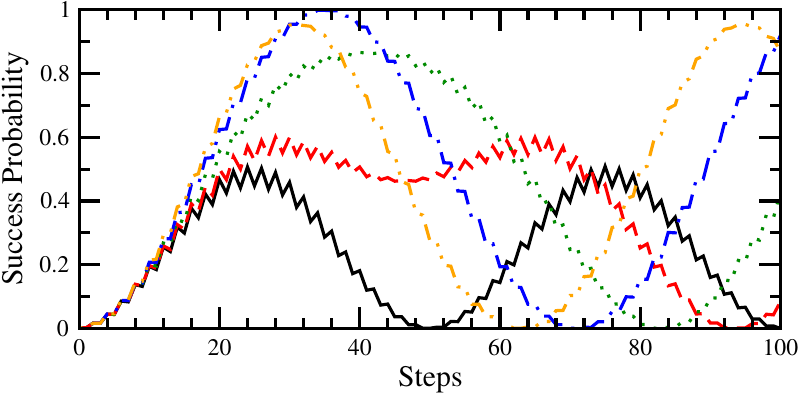}
                \label{fig:johnson_12_4}
        }

	\subfloat[] {
                \includegraphics[height=1.4in]{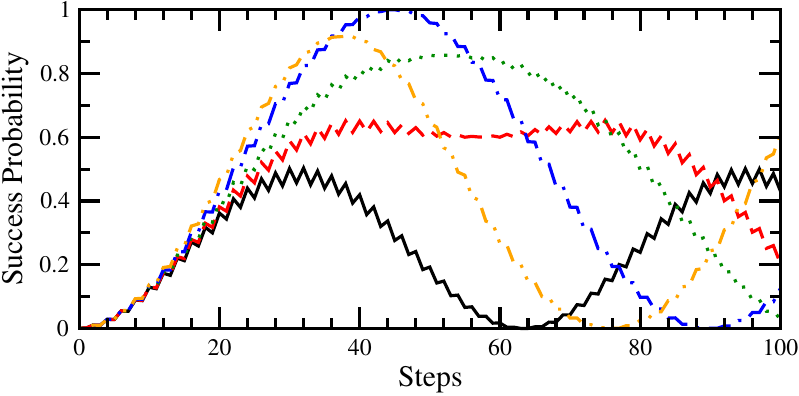}
                \label{fig:johnson_12_5}
        } \quad
	\subfloat[] {
                \includegraphics[height=1.4in]{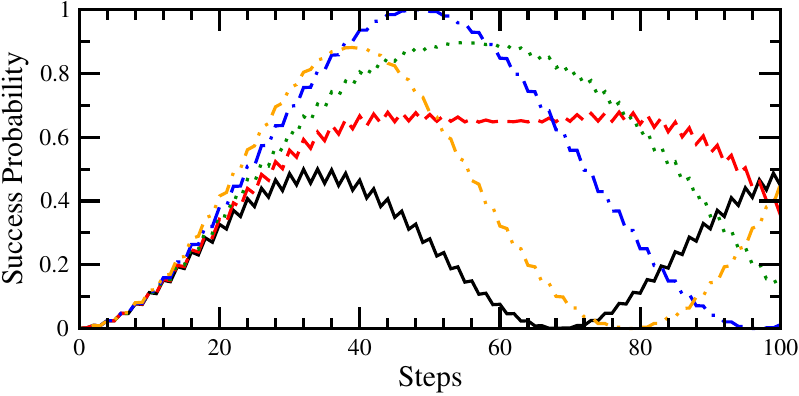}
                \label{fig:johnson_12_6}
        }

	\caption{(a) The Johnson graph $J(5,3)$. (b) Lackadaisical quantum search on $J(12,4)$, which has $495$ vertices and degree $32$. The solid black curve is $\ell = 0$, dashed red is $0.01$, dotted green is $0.03$, the dot-dashed blue is $0.064646$, and the dot-dot-dashed orange is $0.1$. (c) Lackadaisical quantum search on $J(12,5)$, which has $792$ vertices and degree $35$. The solid black curve is $\ell = 0$, dashed red is $0.01$, dotted green is $0.02$, the dot-dashed blue is $0.044192$, and the dot-dot-dashed orange is $0.08$. (d) Lackadaisical quantum search on $J(12,6)$, which has $924$ vertices and degree $36$. The solid black curve is $\ell = 0$, dashed red is $0.01$, dotted green is $0.02$, the dot-dashed blue is $0.038961$, and the dot-dot-dashed orange is $0.08$.} 
\end{center}
\end{figure*}

Johnson graphs are another notable family of vertex-transitive graphs. They have a history in quantum computing as the underlying graphs supporting the quantum-walk algorithms for element distinctness \cite{Ambainis2004} and verifying matrix products \cite{Buhrman2006}. The vertices of the Johnson graph $J(n,k)$ are the $k$-element subsets of $n$ symbols, and vertices are adjacent if they differ in exactly one symbol. For example, \fref{fig:johnson_graph} shows $J(5,3)$. The $n$ symbols are $a,b,c,d,e$, and the vertices are the $3$-element subsets $abc$, $abd$, $abe$, $acd$, $ace$, $ade$, $bcd$, $bce$, $bde$, and $cde$. Vertices are adjacent if they differ in exactly one symbol, or equivalently, if they share exactly two symbols. In general, the number of vertices is the combination ``$n$ choose $k$,'' and the degree is $k(n-k)$. Since the complete graph of $N$ vertices (cf.~\fref{fig:complete}) is the Johnson graph $J(N,1)$, and the triangular graph $T_m$ (cf.~\fref{fig:srg_triangular}) is the Johnson graph $J(m,2)$, this raises the question of whether other Johnson graphs follow the observation that $\ell = d/N$ optimally boosts the success probability.

Figure~\ref{fig:johnson_12_4} shows search using the lackadaisical quantum walk on $J(12,4)$, which has $N = 495$ vertices and degree $d = 32$. Following our observation, the dot-dashed blue curve shows the success probability reaching $1$ when $\ell = d/N = 32/495 = 0.064646$.

Next, \fref{fig:johnson_12_5} shows search on $J(12,5)$, which has $N = 792$ vertices and degree $d = 35$. Again, the dot-dashed blue curve corresponding to $\ell = d/N = 35/792 = 0.044192$ reaches a peak success probability of $1$.

Finally, \fref{fig:johnson_12_6} is search on $J(12,6)$, which has $N = 924$ vertices and degree $d = 36$. This continues to support our proposal, as $\ell = d/N = 36/924 = 0.038961$ is the dot-dashed blue curve, which reaches a success probability of $1$.


\section{Hypercube}

\begin{figure}
\begin{center}
        \subfloat[] {
                \includegraphics[height=1.4in]{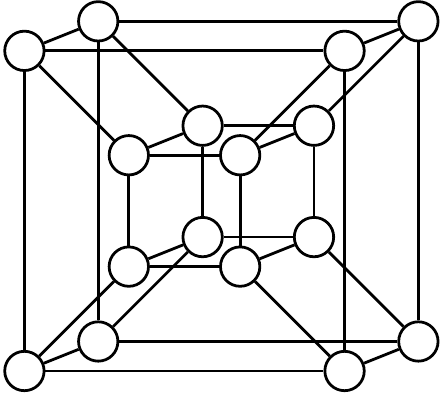}
                \label{fig:hypercube_graph}
        } \quad
	\subfloat[] {
		\includegraphics[height=1.4in]{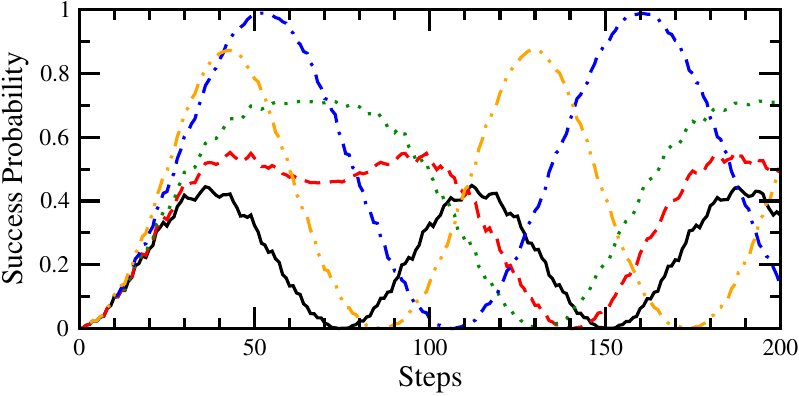}
		\label{fig:hypercube_10}
        }

	\caption{(a) A 4-dimensional hypercube, which has $2^4 = 16$ vertices. (b) Lackadaisical quantum search on the 10-dimensional hypercube, which has $1024$ vertices. The solid black curve is $\ell = 0$, dashed red is $0.0015$, dotted green is $0.003$, the dot-dashed blue is $0.0097656$, and the dot-dot-dashed orange is $0.02$.} 
\end{center}
\end{figure}

Our final investigation is the $n$-dimensional hypercube, which was the first graph on which quantum walks searched \cite{SKW2003}. In 2D, it is a square; in 3D, it is a cube; and in 4D, it is a tesseract, an example of which is shown in \fref{fig:hypercube_graph}. In general, the $n$-dimensional hypercube has $N = 2^n$ vertices, which can be labeled using binary strings of length $n$. They are adjacent if and only if their Hamming distance is $1$, i.e., if they differ in a single bit. Thus, the degree of the hypercube is $d = n = \log_2 N$.

Search on the 10D hypercube is shown in \fref{fig:hypercube_10}. It has $N = 1024$ vertices and degree $d = 10$. In support of our observation, the dot-dashed blue curve corresponding to $\ell = d/N = 10/1024 = 0.0097656$ reaches a success probability of $1$.


\section{Analytical Proof}

One approach to analytically proving these results is using degenerate perturbation theory \cite{Wong5,Wong18}. As a simple example, let us analyze search on the complete graph by lackadaisical quantum walk. This is simple enough that it was analyzed in \cite{Wong10,Wong27} without needing degenerate perturbation theory, but here we rederive the results using degenerate perturbation theory as an illustration. Following \cite{Wong27}, if we label the marked vertex $a$ and the unmarked vertices $b$, then with self-loops, the system evolves in a 4D subspace spanned by
\begin{align*}
    \ket{aa} &= \ket{a} \otimes \ket{a}, \\
    \ket{ab} &= \ket{a} \otimes \frac{1}{\sqrt{N-1}} \sum_{b \in \text{unmarked}} \ket{b}, \\
    \ket{ba} &= \frac{1}{\sqrt{N-1}} \sum_{b \in \text{unmarked}} \ket{b} \otimes \ket{a}, \\
    \ket{bb} &= \frac{1}{\sqrt{N-1}} \sum_{b \in \text{unmarked}} \ket{b} \otimes \frac{1}{\sqrt{N+\ell-2}} \left( \sqrt{\ell} \ket{b} + \sum_{\substack{b' \in \text{unmarked} \\ b' \ne b}} \ket{b'} \right).
\end{align*}
Then from \cite{Wong27}, the initial state \eqref{eq:psi0} in this basis is
\begin{eqnarray*}
	\ket{\psi(0)} 
		= \frac{1}{\sqrt{N(N+\ell-1)}} \Big[ &\sqrt{\ell} \ket{aa} +\sqrt{N-1} \ket{ab} + \sqrt{N-1} \ket{ba} \\
		&+ \sqrt{(N-1)(N+\ell-2)} \ket{bb} \Big].
\end{eqnarray*}
Note for large $N$, this initial state is approximately $\ket{bb}$. Next, if we represent the 4D basis states as $\ket{aa} = \begin{pmatrix} 1 & 0 & 0 & 0 \end{pmatrix}^\intercal, \ket{ab} = \begin{pmatrix} 0 & 1 & 0 & 0 \end{pmatrix}^\intercal, \dots, \ket{bb} = \begin{pmatrix} 0 & 0 & 0 & 1 \end{pmatrix}^\intercal$, then from \cite{Wong27}, the search operator \eqref{eq:U} in this basis is
\begin{equation*}
	U = \left( \!\! \begin{array}{cccc}
		\cos\theta & -\sin\theta & 0 & 0 \\
		0 & 0 & -\cos\phi & \sin\phi \\
		-\sin\theta & -\cos\theta & 0 & 0 \\
		0 & 0 & \sin\phi & \cos\phi \\
	\end{array} \!\! \right),
\end{equation*}
where
\[ \cos\theta = \frac{N-\ell-1}{N+\ell-1}, \quad {\rm and} \quad \sin\theta = \frac{2\sqrt{\ell(N-1)}}{N+\ell-1}, \]
and
\[ \cos\phi = \frac{N+\ell-3}{N+\ell-1}, \quad {\rm and} \quad \sin\phi = \frac{2\sqrt{N+\ell-2}}{N+\ell-1}. \]
For large $N$, we can split $U$ into its leading- and higher-order terms:
\[ U = \underbrace{\begin{pmatrix}
    1 & 0 & 0 & 0 \\
    0 & 0 & -1 & 0 \\
    0 & -1 & 0 & 0 \\
    0 & 0 & 0 & 1 \\
\end{pmatrix}}_{U_0} + \underbrace{\begin{pmatrix}
    0 & -2\sqrt{\frac{\ell}{N}} & 0 & 0 \\
    0 & 0 & 0 & \frac{2}{\sqrt{N}} \\
    -2\sqrt{\frac{\ell}{N}} & 0 & 0 & 0 \\
    0 & 0 & \frac{2}{\sqrt{N}} & 0 \\
\end{pmatrix}}_{U_1} + O\left( \frac{1}{N} \right). \]
Then, to constant-order, the eigenvectors and eigenvalues of $U$ are the eigenvectors and eigenvalues of $U_0$, which are
\begin{align*}
    &\ket{aa}, \quad 1, \\
    &\ket{bb}, \quad 1, \\
    &\ket{-} = \frac{1}{\sqrt{2}} \left( \ket{ab} - \ket{ba} \right), \quad 1, \\
    &\ket{+} = \frac{1}{\sqrt{2}} \left( \ket{ab} + \ket{ba} \right), \quad -1.
\end{align*}
Using these eigenvectors of $U_0$, we can find approximate eigenvectors of $U_0 + U_1$ using degenerate perturbation theory. Since $\ket{aa}$, $\ket{bb}$, and $\ket{-}$ are degenerate eigenvectors of $U_0$, three superpositions of them are eigenvectors of $U_0 + U_1$ for large $N$. That is, for large $N$, three of the four eigenvectors of $U_0 + U_1$ take the form
\[ \alpha_a \ket{aa} + \alpha_b \ket{bb} + \alpha_- \ket{-}, \]
where the coefficients $\alpha_a$, $\alpha_b$, and $\alpha_-$ and the eigenvalues $\lambda$ can be found by solving
\[ \begin{pmatrix}
    U_{aa} & U_{ab} & U_{a-} \\
    U_{ba} & U_{bb} & U_{b-} \\
    U_{-a} & U_{-b} & U_{--} \\
\end{pmatrix} \begin{pmatrix}
    \alpha_a \\
    \alpha_b \\
    \alpha_- \\
\end{pmatrix} = \lambda \begin{pmatrix}
    \alpha_a \\
    \alpha_b \\
    \alpha_- \\
\end{pmatrix}, \]
where $U_{ab} = \langle aa | (U_0 + U_1) | bb \rangle$, etc. Evaluating the matrix elements,
\[ \begin{pmatrix}
    1 & 0 & -\sqrt{\frac{2\ell}{N}} \\
    0 & 1 & -\sqrt{\frac{2}{N}} \\
    \sqrt{\frac{2\ell}{N}} & \sqrt{\frac{2}{N}} & 1 \\
\end{pmatrix} \begin{pmatrix}
    \alpha_a \\
    \alpha_b \\
    \alpha_- \\
\end{pmatrix} = \lambda \begin{pmatrix}
    \alpha_a \\
    \alpha_b \\
    \alpha_- \\
\end{pmatrix}. \]
Solving this eigenvalue relation, three approximate eigenvectors and eigenvalues of $U_0 + U_1$ are 
\begin{align*}
    & u = i \sqrt{\frac{\ell}{\ell+1}} \ket{aa} + \frac{i}{\sqrt{\ell+1}} \ket{bb} + \ket{-}, \quad 1 + i \sqrt{\frac{2(\ell+1)}{N}} \approx e^{i\sigma}, \\
    & v = -i \sqrt{\frac{\ell}{\ell+1}} \ket{aa} - \frac{i}{\sqrt{\ell+1}} \ket{bb} + \ket{-}, \quad 1 - i \sqrt{\frac{2(\ell+1)}{N}} \approx e^{-i\sigma}, \\
    & w = -\frac{1}{\sqrt{\ell}} \ket{aa} + \ket{bb}, \quad 1,
\end{align*}
where $\sigma \approx \sqrt{2(\ell+1)/N}$. Note $u$, $v$, and $w$ are unnormalized, but they can be easily normalized if desired. The initial state can be expressed in terms of these approximate eigenvectors:
\[ \ket{\psi(0)} \approx \ket{bb} = \frac{-i}{2\sqrt{\ell+1}} \left( u - v + i\frac{2\ell}{\sqrt{\ell+1}} w \right). \]
Applying the search operator $t_*$ times, where
\[ t_* = \frac{\pi}{\sigma} = \frac{\pi}{\sqrt{2(\ell+1)}} \sqrt{N}, \]
the system evolves to
\begin{align*}
    \ket{\psi(t)} 
        &\approx \frac{-i}{2\sqrt{\ell+1}} \left( u e^{i\pi} - v e^{-i\pi} + i\frac{2\ell}{\sqrt{\ell+1}} w \right) \\
        &\approx \frac{-i}{2\sqrt{\ell+1}} \left( -u + v + i\frac{2\ell}{\sqrt{\ell+1}} w \right) \\
        &= \frac{-2\sqrt{\ell}}{\ell+1} \ket{aa} + \frac{\ell-1}{\ell+1} \ket{bb}.
\end{align*}
Taking the square of the amplitude of $\ket{aa}$, the success probability at time $t_*$ is
\[ p_* = \frac{4\ell}{(\ell+1)^2}. \]
This analytic result, which we proved using degenerate perturbation theory, is in agreement with the non-perturbative calculation in \cite{Wong10,Wong27}. Then, when $\ell = 1$, the success probability reaches $1$ at time $\pi\sqrt{N}/2$, as seen in \fref{fig:complete_1024}.

This proof method can be similarly applied to each family of graphs that we simulated. For complete bipartite graphs, this was done in \cite{Wong32}. For strongly regular graphs, there are only three kinds of vertices: the marked vertex, vertices adjacent to the marked vertex, and vertices non-adjacent to the marked vertex \cite{Wong5}. Including self-loops and the coin degree of freedom, the system evolves in a 7D subspace, regardless of the parameters $(N,k,\lambda,\mu)$. Depending on how the parameters scale with $N$, the leading- and higher-order terms would change \cite{Wong5}. Johnson graphs $J(n,k)$ are distance transitive, so for fixed $k$, the system evolves in a fixed-dimensional subspace \cite{Wong20}. The perturbative calculation would have to be repeated for each value of $k$, however, and the countably infinite number of values of $k$ makes this calculation impractical. Lattices and the hypercube evolve in subspaces whose dimensions increase with $N$, but degenerate perturbation theory has been shown to successfully handle such cases, at least for continuous-time quantum walks \cite{Wong8}.

Since a different perturbative calculation could be required for each family of graphs, degenerate perturbation theory seems like an unnatural approach to prove that $\ell = d/N$ is optimal for all vertex-transitive graphs in general. Thankfully, a general proof of our observation is forthcoming by H{\o}yer and Yu \cite{Hoyer2020}, who frames the lackadaisical quantum walk as an interpolated walk that is quantized using Szegedy's correspondence \cite{Szegedy2004}.


\section{Conclusion}

The lackadaisical quantum walk has been explored as a tool for speeding up spatial search. We observed that the existing results for search on vertex-transitive graphs with a unique marked vertex followed a trend that a self-loop weight of $\ell = d/N$ boosted the success probability to 1, or in the case of the torus and cycle, boosted it to a constant, which is the best possible scaling for the success probability. Investigating this trend, we provided numerical results for search on arbitrary-dimensional cubic lattices with periodic boundary conditions, families of strongly regular graphs, Johnson graphs, and the hypercube, and they all followed the trend that $\ell = d/N$ was optimal. From this, we propose that for search on vertex-transitive graphs with a unique marked vertex, $\ell = d/N$ is optimal. While degenerate perturbation theory can be used to prove this for specific families of graphs, it seems to be a less than ideal proof method for this general observation. A forthcoming general proof by H{\o}yer and Yu, however, overcomes this difficulty by framing the lackadaisical quantum walk as an interpolated walk.


\begin{acknowledgements}
	Thanks to Peter H{\o}yer and Zhan Yu for useful discussions. This work was partially supported by T.W.'s startup funds from Creighton University.
\end{acknowledgements}


\bibliographystyle{qinp}
\bibliography{refs}

\end{document}